\documentclass[aps,pra,twocolumn,showpacs,superscriptaddress,preprintnumbers,amsmath,amssymb,nofootinbib]{revtex4}

\usepackage{graphicx}
\usepackage{mathtools}
\usepackage{gensymb}
\begin{document}

% Use the \preprint command to place your local institutional report
% number in the upper righthand corner of the title page in preprint mode.
% Multiple \preprint commands are allowed.
% Use the 'preprintnumbers' class option to override journal defaults
% to display numbers if necessary
%\preprint{}

%Title of paper
\title{Modification of incoherent scattering under the influence of system eigenmodes}

% repeat the \author .. \affiliation  etc. as needed
% \email, \thanks, \homepage, \altaffiliation all apply to the current
% author. Explanatory text should go in the []'s, actual e-mail
% address or url should go in the {}'s for \email and \homepage.
% Please use the appropriate macro foreach each type of information
% \affiliation command applies to all authors since the last
% \affiliation command. The \affiliation command should follow the
% other information
% \affiliation can be followed by \email, \homepage, \thanks as well.

\author{R.S. Puzko}
\email[]{roman998@mail.ru}
\affiliation{All-Russia Research Institute of Automatics, 22, ul. Sushchevskaya, Moscow 127055, Russia}
\affiliation{Moscow Institute of Physics and Technology, 9 Institutskiy per., Dolgoprudny, Moscow Region, 141700, Russia}

\author{A.D. Brozhek}
\affiliation{A.M. Prokhorov General Physics Institute, Russian Academy of Sciences, Vavilov str. 38, Moscow 119991, Russia}

\author{V.I. Fabelinsky}
\affiliation{A.M. Prokhorov General Physics Institute, Russian Academy of Sciences, Vavilov str. 38, Moscow 119991, Russia}

\author{D.N. Kozlov}
\affiliation{Moscow Institute of Physics and Technology, 9 Institutskiy per., Dolgoprudny, Moscow Region, 141700, Russia}
\affiliation{A.M. Prokhorov General Physics Institute, Russian Academy of Sciences, Vavilov str. 38, Moscow 119991, Russia}

\author{Y.N. Polivanov}
\affiliation{A.M. Prokhorov General Physics Institute, Russian Academy of Sciences, Vavilov str. 38, Moscow 119991, Russia}

\author{V.V. Smirnov}
\affiliation{A.M. Prokhorov General Physics Institute, Russian Academy of Sciences, Vavilov str. 38, Moscow 119991, Russia}

\author{A.N. Lagarkov}
\affiliation{Institute for Theoretical and Applied Electromagnetics Russian Academy of Sciences, 13, ul. Izhorskaya, Moscow 125412, Russia}

\author{A.K. Sarychev}
\affiliation{Institute for Theoretical and Applied Electromagnetics Russian Academy of Sciences, 13, ul. Izhorskaya, Moscow 125412, Russia}

\author{K.N. Afanasiev}
\affiliation{Institute for Theoretical and Applied Electromagnetics Russian Academy of Sciences, 13, ul. Izhorskaya, Moscow 125412, Russia}

\author{I.A. Ryzhikov}
\affiliation{Institute for Theoretical and Applied Electromagnetics Russian Academy of Sciences, 13, ul. Izhorskaya, Moscow 125412, Russia}
\affiliation{Bauman Moscow State Technical University, 2nd Baumanskaya str., 5/1, Moscow 105005, Russia}
\affiliation{Emanuel Institute of Biochemical Physics, Russian Academy of Science, Kosygina str., 4, Moscow 119334, Russia}

\author{I.A. Boginskaya}
\affiliation{Institute for Theoretical and Applied Electromagnetics Russian Academy of Sciences, 13, ul. Izhorskaya, Moscow 125412, Russia}

\author{M.V. Sedova}
\affiliation{Institute for Theoretical and Applied Electromagnetics Russian Academy of Sciences, 13, ul. Izhorskaya, Moscow 125412, Russia}

\author{I.N. Kurochkin}
\affiliation{Emanuel Institute of Biochemical Physics, Russian Academy of Science, Kosygina str., 4, Moscow 119334, Russia}

\author{A.M. Merzlikin}
\email[]{merzlikin_a@mail.ru}
\affiliation{All-Russia Research Institute of Automatics, 22, ul. Sushchevskaya, Moscow 127055, Russia}
\affiliation{Moscow Institute of Physics and Technology, 9 Institutskiy per., Dolgoprudny, Moscow Region, 141700, Russia}
\affiliation{Institute for Theoretical and Applied Electromagnetics Russian Academy of Sciences, 13, ul. Izhorskaya, Moscow 125412, Russia}

\date{\today}
	
\begin{abstract}
The optical properties of inhomogeneous cerium dioxide ($CeO_2$) films on aluminum ($Al$) sublayer are investigated. The dependencies of the reflection, scattering, and absorption of coherent electromagnetic radiation as functions of the incidence angle and polarization are studied. The experimental and numerical studies show the existence of an incidence angle at which the scattering and absorption of coherent radiation with s-polarization increase simultaneously and the specular reflection becomes just about a few percents. The angle corresponds to an excitation of Fabry-Perot resonator modes. This effect opens up great prospects for manipulation (tuning) of the reflection from the inhomogeneous films.
\end{abstract}
	
% insert suggested PACS numbers in braces on next line
\pacs{???}
% insert suggested keywords - APS authors don't need to do this
%\keywords{}
	
%\maketitle must follow title, authors, abstract, \pacs, and \keywords
\maketitle
	
% body of paper here - Use proper section commands
% References should be done using the \cite, \ref, and \label commands
\section{Introduction}
The scattering of light by inhomogeneities is of interest for various fields of physics and engineering, ranging from astronomical and atmospheric studies \cite{IntroAtm,IntroAtm1} to applications in medicine and technology (increasing the efficiency of solar cells \cite{IntroSolarCells}, analyzing biological samples \cite{IntroBio}, etc.).

Among the light scattering problems, two important cases are widely discussed in literature: volume scattering in an inhomogeneous medium \cite{VolumeGarciaSiewert,VolumeGarcia,VolumeAduev} and scattering by the roughness of the sample boundaries \cite{SurfaceGarciaLlamos,SurfaceMcGurn,SurfaceVisperinas,SurfaceSGil}.

In the case of volume scattering, the propagation of an electromagnetic wave through a medium containing a certain volume concentration of inhomogeneities is considered \cite{VolumeIntroGarcia,VolumeIntroZvekov} (a composite material with inclusions). The processes of scattering and absorption are described in the diffusion approximation \cite{Sheng,Ishimaru} by the Kubelka-Munk theory \cite{Ishimaru} and using the Monte Carlo method \cite{MonteCarlo1,MonteCarlo2}. Particular attention is paid to light scattering inside the bounded structures where it is possible to identify regions with different optical properties; e.g. light scattering inside multilayer systems is considered in a number of papers \cite{VolumeBoundary0,VolumeIntroZvekov,VolumeBoundary1,VolumeBoundary2}.

Another class of light scattering problems devoted to rough surfaces has been studied in detail by many authors \cite{Ogura1,Ogura2,SanchezGil,SurfaceVisperinas,NietoVisperinas,SurfaceMcGurn,SurfaceSGil,Perturbation1,Perturbation2,Perturbation3,Perturbation4}. As a rule, the surfaces having small height fluctuations (in comparison with the illumination wavelength) are considered within the frameworks of theoretical works. This allows one to construct perturbation theory \cite{SurfaceSGil,Perturbation3} or an equivalent method of stochastic functionals \cite{Ogura1,Ogura2}. These approaches were used in \cite{SurfaceSGil,Ogura2} to study single and double scattering processes in detail (in particular, the effects of backward coherent scattering and satellite peaks are considered). It is worth noting here that two problems can be distinguished: light scattering by the roughness of half-space boundary \cite{Perturbation4} and light scattering by rough boundaries of structured samples, in particular, layered systems \cite{Ogura2,Perturbation1,Perturbation3}. In the latter case, the light scattering is significantly affected by the modes of the system. For example, the effect of resonant scattering was discovered in the layered system supporting the guided modes \cite{SurfaceGarciaLlamos}. The resonant scattering occurs at the angles of guided mode excitation. Experimental investigations of scattering spectra for films with a rough surface are presented in \cite{SurfaceGarciaLlamos,Chaikina,GarciaLlamos}. 

The presence of system eigenmodes may result in significant scattering features. It should be noted, in this connection, that the high contrast of material dielectric permittivity and the small Ohmic losses result in the resonances with a high Q-factor. Therefore, the films of cerium dioxide—a non-absorbing material with a large refractive index used for single- and multilayer coatings—are of particular interest \cite{CeO21}. Moreover, the microstructure of cerium dioxide layers depends strongly on the manufacturing method of the film and can be completely inhomogeneous \cite{CeO21,CeO20,CeO22,CeO23,CeO24,CeO25,CeO26}. Because of the large refractive index together with an inhomogeneous microstructure, cerium dioxide ($CeO_2$) structures are the attractive objects for light scattering studies.

This paper is devoted to laser light scattering inside CeO2 films produced by electron-beam deposition in vacuum \cite{Kurochkin}. It is worth noting that a CeO2 film structure possesses complex morphology, in particular a faceted one \cite{Kurochkin}. The complex nanostructure of the CeO2 film can significantly affect the optical properties \cite{CeO22,CeO23}. Furthermore, the light scattering by film inhomogeneities may lead to an effective attenuation of the coherent radiation. In the paper we propose a simple model of light scattering by inhomogeneities. The model makes it possible to describe the scattering and reflection features found experimentally.

The paper is organized as follows. Firstly, the investigated samples of cerium oxide and sample morphology are considered (Section 2). Then, the scheme of the experiment on laser light reflection and scattering measurements is presented (Section 3). Section 4 is devoted to the theoretical model describing light scattering. The theoretical consideration starts with a case of weak scattering, when the power scattered by inhomogeneities is negligibly small in comparison with the power of the incident wave. Then, the case of strong scattering is studied and the scattering parameter used in calculations is introduced. Finally, in Section 5 the results are briefly summarized.

\section{Cerium dioxide sample}
The studied sample is a two-layered system placed on a substrate of ceramic alumina ($Al_2O_3$). The substrate was covered by a 100–150 nm thick aluminum ($Al$) sublayer. The $CeO_2$ film was deposited on the $Al$ sublayer. The film thickness was 2100 nm.

A technological route to produce cerium dioxide films consists of the following steps:

- pretreatment of an $Al_2O_3$ substrate surface by washing in isopropyl alcohol and ion cleaning in a vacuum chamber ($10^{-2}$ Torr) by an ion beam (beam current 150 mA, voltage 1.5 kV) for 15 minutes;

- electron-beam deposition of the $Al$ sublayer by electron beam evaporation (electron beam current 100 mA, voltage 8kV);

- placing high purity $CeO_2$ powder into a water-cooled copper crucible, and its electron-beam evaporation (beam current 30 mA, voltage 8 kV).

The deposited film thickness was measured using optical inspection of the interference maxima and minima at 600 nm using the control glass sample.

\begin{figure}
	\begin{minipage}{0.49\linewidth}
		\includegraphics[width=1.0\linewidth]{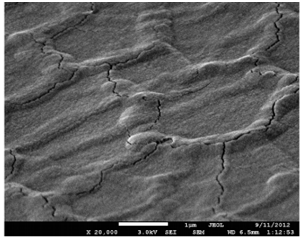}
		(a)
	\end{minipage}
	\begin{minipage}{0.49\linewidth}
		\includegraphics[width=1.0\linewidth]{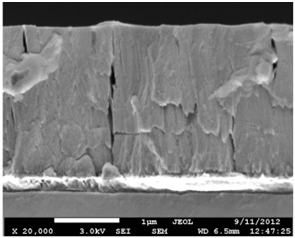}
		(b)
	\end{minipage}
	\caption{a) A SEM image of a CeO2 film; b) a cleavage of a $CeO_2$ film.\label{SEM}}
\end{figure}

An important feature of the system is the structure of the $CeO_2$ layer: firstly, this film has a sufficiently large thickness (up to several wavelengths in the optical spectrum); secondly, an inhomogeneous faceted (islet) structure (see Fig.~\ref{SEM}a) is formed as a result of the manufacturing process. The $CeO_2$ islands, with characteristic transverse dimensions $\sim2-4\, {\mu}\mathrm{m}$, are divided by narrow cracks ($\sim10-70\, \mathrm{nm}$). Moreover, the islands themselves also support the density inhomogeneities.

Figure~\ref{SEM} shows typical faceted structures of the films produced. A detailed study of the faceted film structure was carried out by using atomic force microscopy (AFM) and scanning electron microscopy (SEM).

\section{Experiment}
\begin{figure}
\includegraphics[width=1\linewidth]{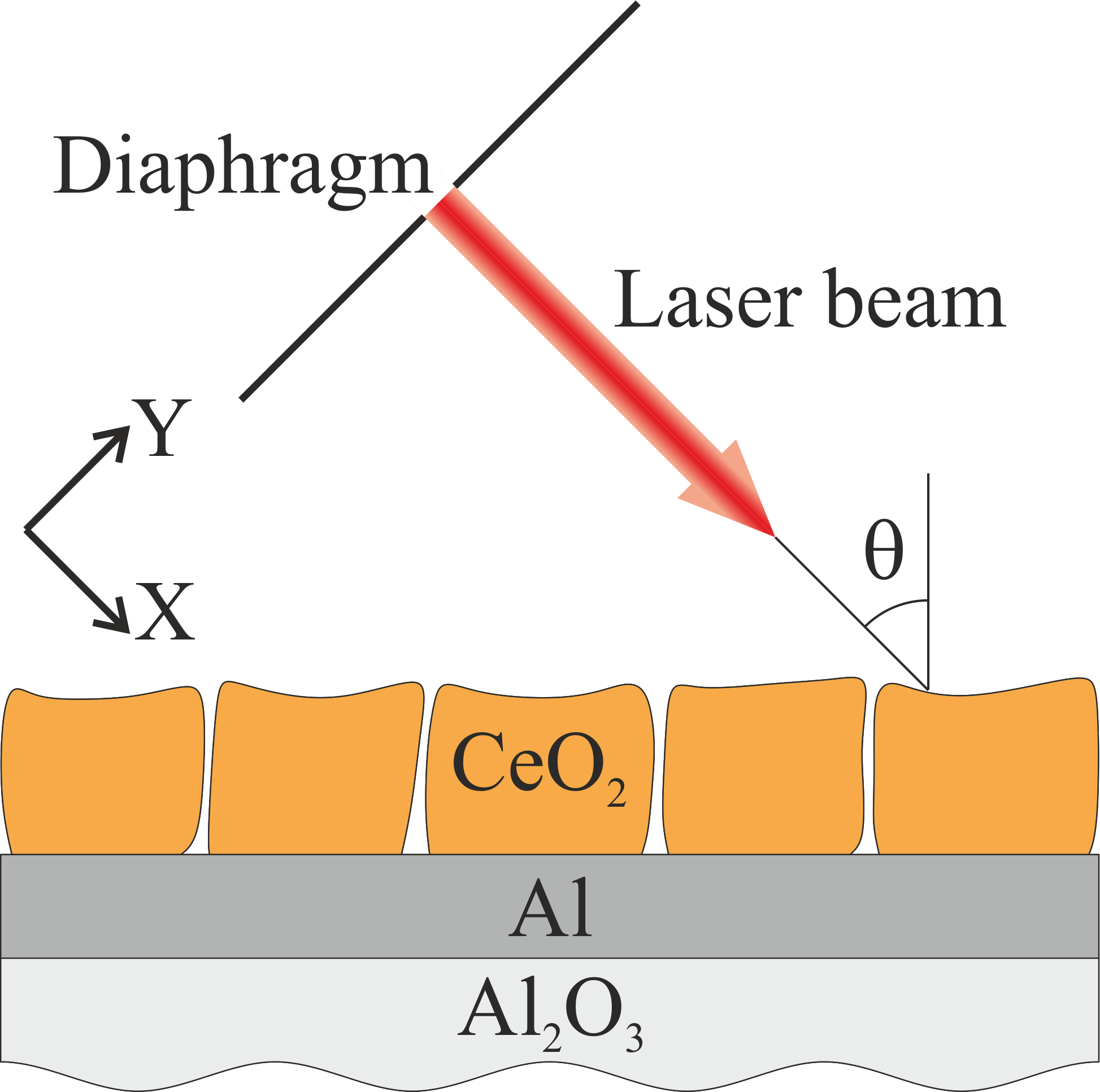}
\caption{The schematic representation of the system under investigation. The laser beam is incident upon the $CeO_2/Al/Al_2O_3$ sample at an angle $\theta$.\label{System}}
\end{figure}

\begin{figure}
	\includegraphics[width=1\linewidth]{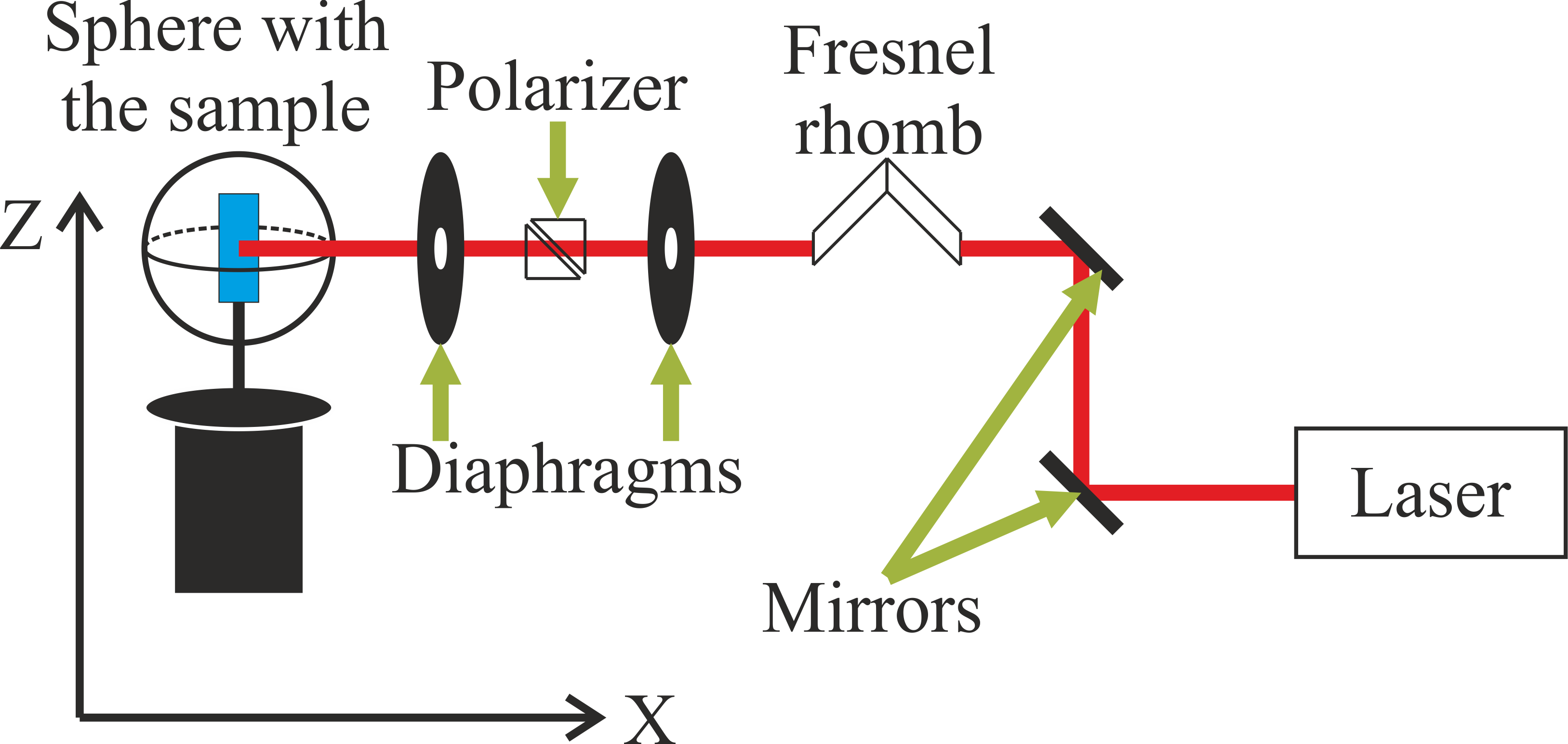}
	\caption{The experimental setup for measuring angular dependences of the total and specular reflection.\label{Setup}}
\end{figure}

The $CeO_2$ film of the two-layered system under investigation possesses the inhomogeneous structure which results in scattering of a coherent light beam. The measurements were carried out to investigate the scattering and reflecting characteristics. Namely, the angular dependence of scattering and specular reflectance on the coherent light incidence angle was measured at a wavelength $\lambda=632.8$ nm (see Fig.~\ref{System}).
The experimental setup for measuring angular dependences of specular ($R$) and total (hemispherical) reflectance (the sum of $R$ and scattering coefficient, $S$) consisted of a He-Ne laser, an integrating sphere, a rotating table with a sample holder, optical elements (mirrors, polarizers, a Fresnel rhomb), and a laser radiation power meter (see Fig.~\ref{Setup}). The measurements were carried out at the wavelength of 632.8 nm with an average laser radiation power of 3–5 mW. The collimated laser beam with a given linear polarization was directed to the sample in a horizontal plane (along the X-axis) through two diaphragms with a diameter of 2 mm (Fig.~\ref{Setup}). The direction of the polarization vector (s- or p-) was established by the double Fresnel rhomb and the Glan-Taylor prism (Polarizer in Fig.~\ref{Setup}).

The total reflectance was measured by using the integrating sphere with a photodetector. To measure the specular reflection, the laser power meter (Coherent Fieldmate) with a silicon photodiode detector (Coherent Op-2VIS) was used. The measurements were performed in the range $4-56\degree$ of the incidence angles $\theta$ adjusted with an accuracy of $1\degree$.

\begin{figure}
	\includegraphics[width=1\linewidth]{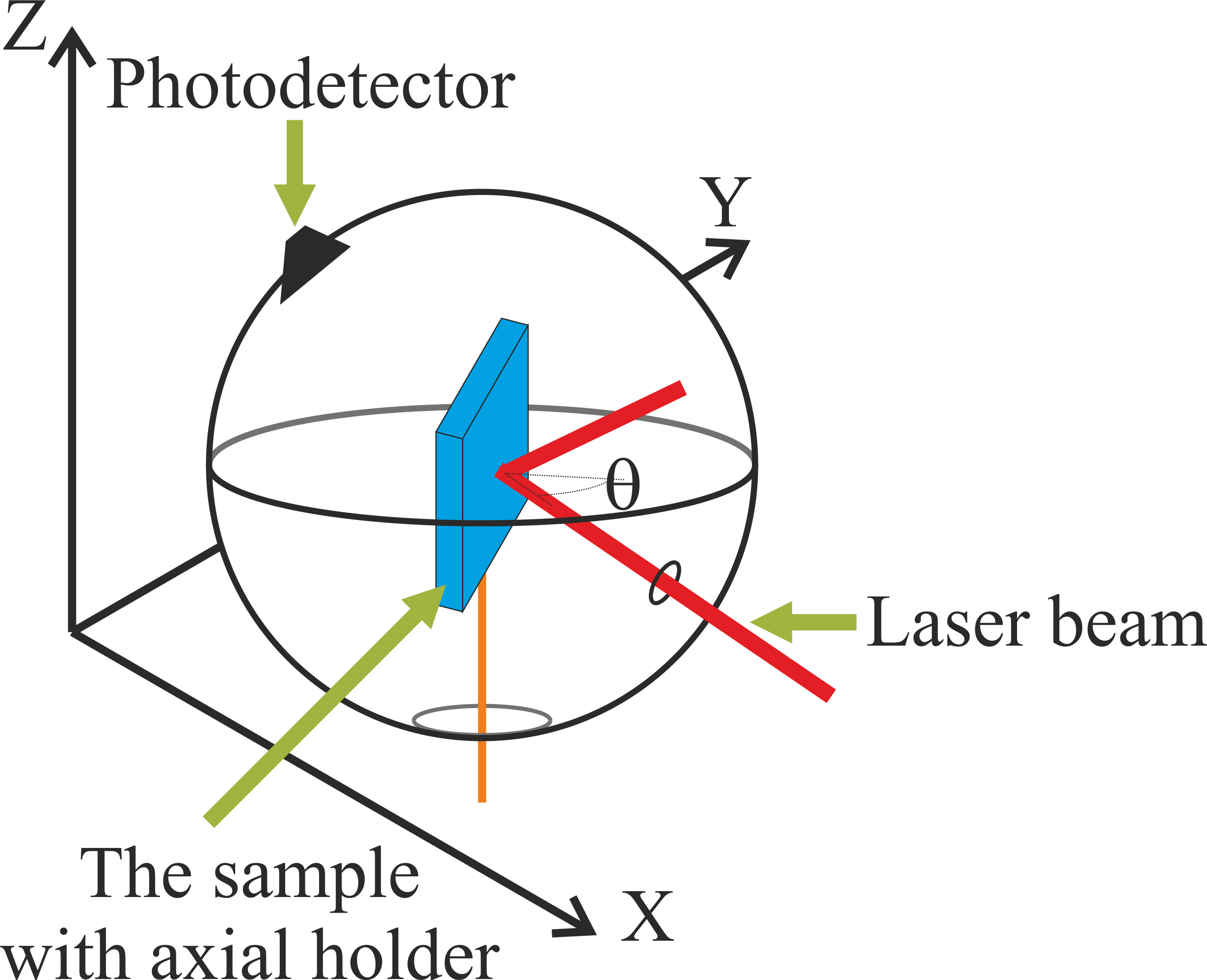}
	\caption{Scheme of the total reflectance measurements by using the integrating sphere.\label{Sphere}}
\end{figure}

The detailed schematic and geometry of the total laser radiation reflectance (R + S) measurements are presented in Fig.~\ref{Sphere}.

The integrating sphere used in measurements was a thin-walled hollow sphere with a diameter of 150 mm. The inner surface of the sphere was covered by powder of barium oxide crystals $BaO$, forming a rough white coating. The coating provided the scattering of light inside the sphere in a wide spectral range (400-650 nm). The sphere had two apertures: a small one, 5 mm in diameter (on the side in Fig.~\ref{Sphere}), and a large one, with a diameter of 25 mm (on the bottom in Fig.~\ref{Sphere}). A large aperture was used to mount a $CeO_2/Al/Al_2O_3$ sample inside the sphere. The sample plate was installed vertically on a thin axial holder, which passed along the vertical Z axis through the center of the base rotating in the horizontal plane (XY). A collimated beam of laser radiation was guided through a small aperture inside the sphere, where it illuminated the sample at a certain angle $\theta$ in the plane of incidence (XY). The radiation specularly reflected from the sample and scattered by it illuminating the inner surface of the sphere and was repeatedly scattered by its rough surface. As a result, the radiation scattered by the sphere’s inner surface fell on the photodetector. The photodetector consisted of a set of multidirectional fibers attached to a silicon photodiode, which registered light in the range 320–1100 nm.

Mounting of the sample on the thin axis prevented significant absorption of light scattered at the sphere surface by the elements of the holder. To reduce the scattered radiation losses through the large aperture, it was covered with a white scattering screen having a small hole in the center for the sample rotation.

The calibration of the integrating sphere photodetector was carried out using a small high-reflectance aluminum mirror, which was mounted instead of the sample. In the process of the calibration, the power of the specularly reflected laser beam was measured by the photodetector and the power meter.

The angular dependence of the specular reflection was measured similarly but without the integrating sphere. In this case, the photodetector of the power meter was being moved in the XY plane around the vertical axis Z to confront its input aperture with the specularly reflected beam.
\begin{figure}
	\includegraphics[width=1\linewidth]{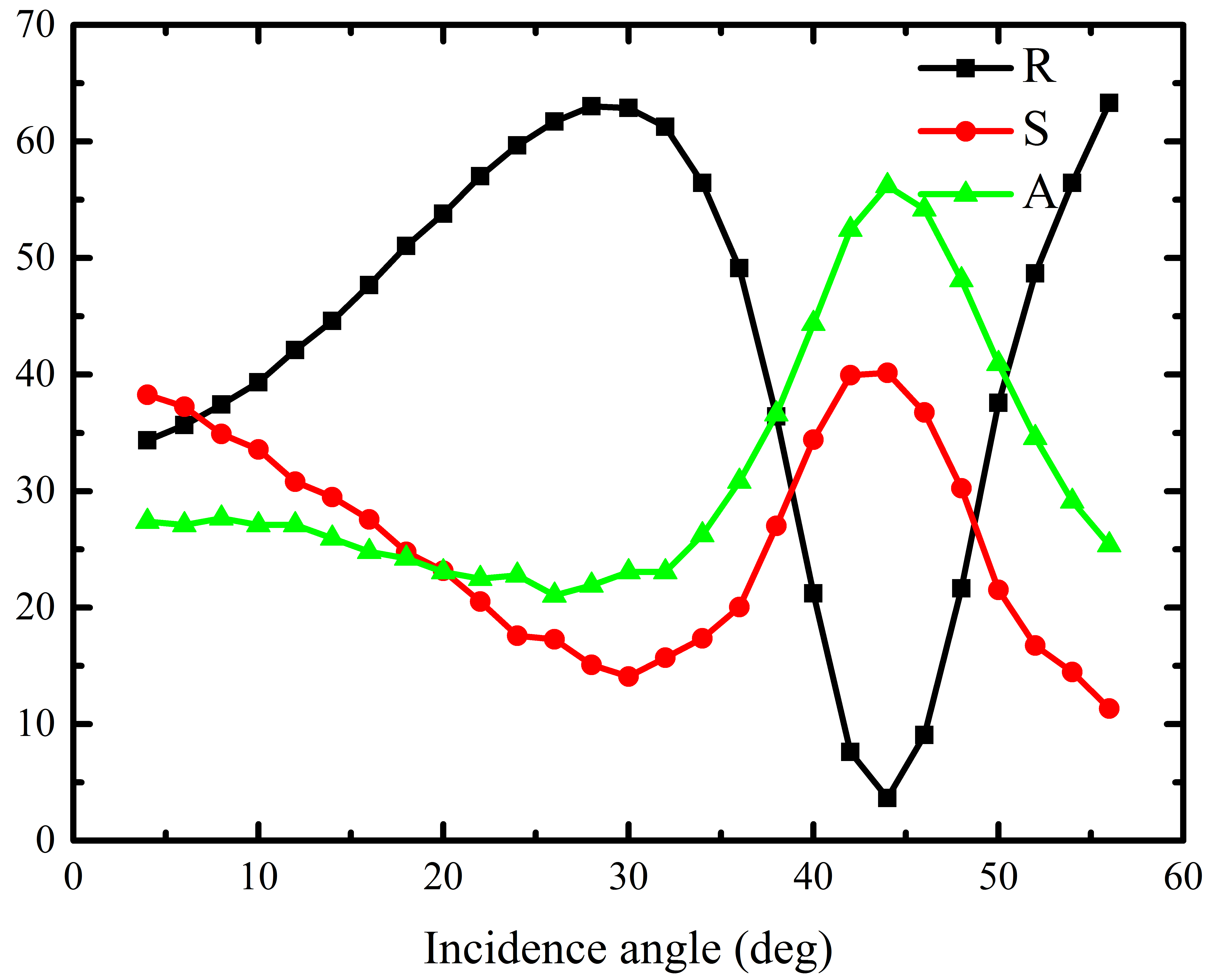}
	\caption{Specular reflection, $R$ (black), scattering, $S$ (red), and absorption, $A$ (green) coefficients of the $CeO_2/Al/Al_2O_3$ sample as a function of the angle of incidence of an s-polarized 633-nm coherent beam.\label{Results}}
\end{figure}

As a result, the dependences of the scattering coefficient and specular reflection on the angle of incidence of an s-polarized light have been derived. The absorption coefficient has been calculated using the energy conservation law $A(\theta)+R(\theta)+S(\theta)=1$. It should be noted that there exist the angles of incidence at which scattering is increased. Absorption increases at the same angles of incidence (see Fig.~\ref{Results}).

\section{Light scattering by the volume inhomogeneities}
To explain the features of the coherent light scattering observed in the experiment, we consider a simple model of scattering in the dipole approximation: the incoherent radiation emitted by the $CeO_2$ film inhomogeneity is considered as the emission of a dipole induced by a coherent external field. As a result of the coherent radiation scattering, part of the coherent beam energy transforms into incoherent scattered radiation energy. Therefore, the coherent part of the radiation decreases as the beam propagates through the material. The radiation scattered by different dipoles will be considered incoherent. In the calculations, we assume that the dipoles representing inhomogeneities (the volume fluctuations of the refractive index) are uniformly distributed within the film volume.

The important simplification used in the paper is considering only single scattering processes and neglecting multiple scattering ones. This approximation is valid for the propagation length of radiation inside the system smaller than the mean free path.

First, we consider a \textbf{model system} in which the scattering is weak and the light attenuation on the layer thickness is insignificant\footnote{This approximation is justified if the scattering cross section of inhomogeneities and their concentration are small, and the film thickness does not exceed several wavelengths (that is, this approximation is correct if the scattered energy is much less than the Ohmic losses in the system).}. We note at once that this approximation does not apply to the $CeO_2$ film (as soon as according to the experimental results the losses due to scattering are comparable with the Ohmic losses (see Fig.~\ref{Results})), and it is considered only for a qualitative explanation of the simultaneous increase in scattering and absorption. Then we proceed to the case of strong scattering for a direct explanation of the experiment.

\subsection{Weak scattering case}
In this section it is considered that the scattering is weak and the energy of the incident coherent radiation does not decrease due to scattering, that is, all the dipoles are in the field of such an external wave as they would be in the absence of scattering.

In the long-wavelength approximation, a dipole moment of a single scatterer is directly proportional to the local electric field. To calculate the scattered field, it is convenient in the given geometry to proceed to the plane wave basis by expanding the dipole radiation into plane waves with different propagation directions and polarizations. Then, the electromagnetic field induced by a dipole in the system (and the scattered radiation field) can be found by considering the propagation of each plane wave of the expansion through the system. The calculation of the electric field by this method is equivalent to the introduction of the Green function for the dipole inside the structure. A detailed description of the calculation procedure is given in the Appendix.

In the weak scattering approximation, the contribution of the scattered radiation to the total energy of the radiation can be neglected, and the sum of the absorption coefficient $A$ and the specular reflection $R$ from the sample is 1
\begin{equation}
R+A=1\label{ConservationLaw}
\end{equation}
i.e. the energy of the scattered radiation is much smaller than that of the coherent radiation. It should be noted that the equation (\ref{ConservationLaw}) expressing the energy conservation law does not include the transmittance. The point here is that the transmittance of the considered system is negligible due to the $Al$ sublayer having a large negative dielectric permittivity.

\begin{figure}
	\includegraphics[width=1\linewidth]{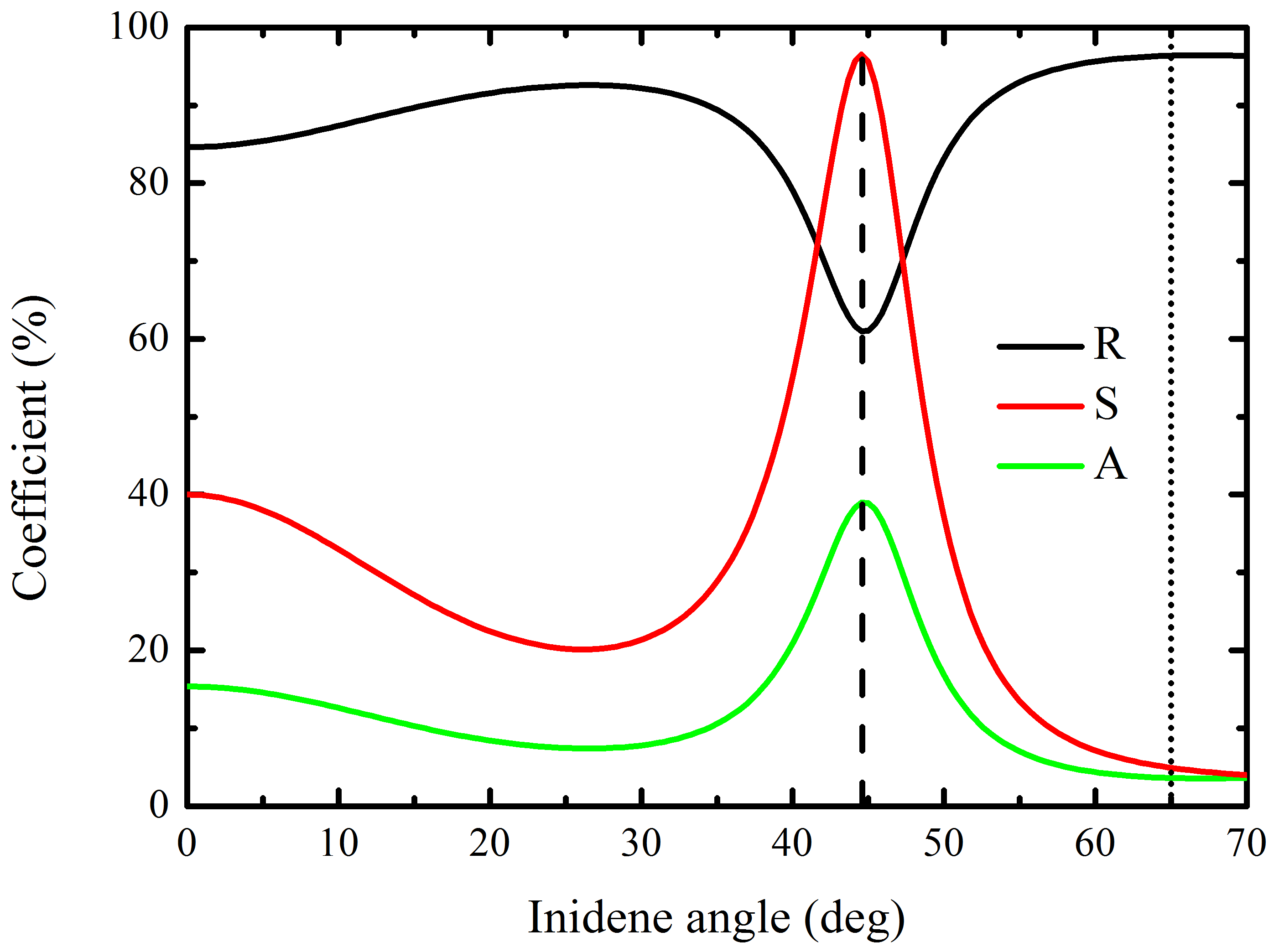}
	\caption{Dependences of the specular reflection, total scattering and absorption coefficients on the incidence angle of s-polarized radiation. The thickness of the $CeO_2$ film - 2096~nm, the dielectric permittivity at the wavelength of 632.8~nm $\varepsilon_{CeO_2}=4.595$ ($n_{CeO_2}=2.1436$). The parameters of the $Al$ sublayer: $d_{Al}=200$~nm, $\varepsilon_{Al}=-51.4+18.4i$~\cite{Rakic}. The dielectric permittivity of the substrate is $\varepsilon_{Al_2O_3}=2.89$ ($n_{Al_2O_3}=1.70$). The total scattering coefficient dependence is normalized to the experimental result at the incidence angle $4\degree$. The vertical dashed line indicates the angle of mode excitation (see Equation 2), and the dotted line corresponds to the Brewster angle.\label{WeakScattering}}
\end{figure}

Figure~\ref{WeakScattering} displays the calculated dependences of the specular reflection, absorption and scattering coefficients of s-polarized radiation on the incidence angle. The calculated values of the reflection and absorption coefficients correspond to the system without defects and density fluctuations, and the transformation of the coherent radiation energy to that of the incoherent one is not taken into account (it is assumed that the latter is infinitesimally small). The scattering coefficient curve is normalized for clarity to the experimental value of the quantity at the incidence angle $4\degree$.

To clarify the presence of the scattering and absorption peaks in Figure~\ref{WeakScattering}, the eigenmodes of the considered system should be taken into account. One can distinguish unbound (radiation) and localized (guided) modes of a dielectric layer.

The localized modes are the guided modes of a one-dimensional waveguide, which are localized inside the layer due to total internal reflection from the boundaries of the layer. In the case, the mode propagates through the layer and loses energy only due to the absorption in the materials.

\begin{figure}
	\includegraphics[width=1\linewidth]{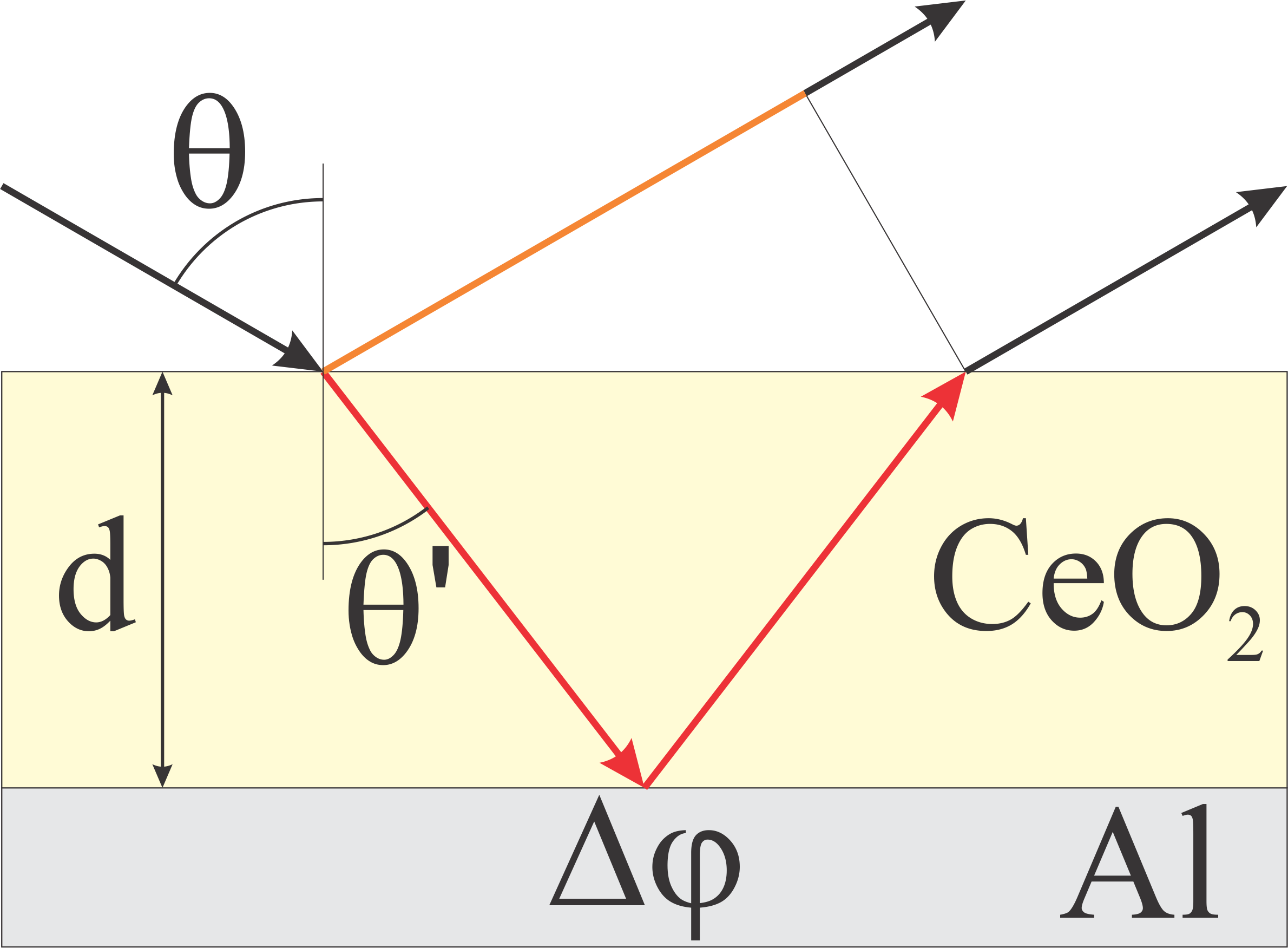}
	\caption{A dielectric waveguide ($CeO_2$) placed on a sublayer of an metal ($Al$).\label{Waveguide}}
\end{figure}

The radiation modes of the waveguide are unbound modes which are not confined to the waveguide. In contrast to the guided modes, these modes have losses caused by the light transmission through the boundaries of the system (see Fig.~\ref{Waveguide}). The particular interest is associated with the resonant radiation modes (RRM) of the waveguide analogous to a Fabry-Perot resonator mode.

The RRM excitation condition is the phase synchronization matching (see Fig.~\ref{Waveguide}). Taking into account the additional phase shift upon reflection from the $Al$ sublayer, the RRM excitation condition can be written as
\begin{equation}
2dnk_0\cos(\theta')+\bigtriangleup\varphi=2\pi m\label{RRM}
\end{equation}
where
\begin{equation}
\bigtriangleup\varphi=\textrm{arg}(r_{CeO_2/Al})=\textrm{arg}\left(\frac{Y_{CeO2}-Y_{Al}}{Y_{CeO2}+Y_{Al}}\right)\label{deltaphi}
\end{equation}
is a phase jump at the boundary $CeO_2/Al$. Here, $r_{CeO_2/Al}$ is the reflection coefficient of light incident from the half-space of $CeO_2$ onto the aluminum half-space, $Y_{CeO_2}$ and $Y_{Al}$ are the angular admittances for the corresponding materials~\cite{Yeh}. It is worth mentioning that condition~(\ref{RRM}) is essentially a condition of the Salisbury screen~\cite{Salisbury}, which indicates that there is a large absorption in the system.
\begin{figure}
	\includegraphics[width=1\linewidth]{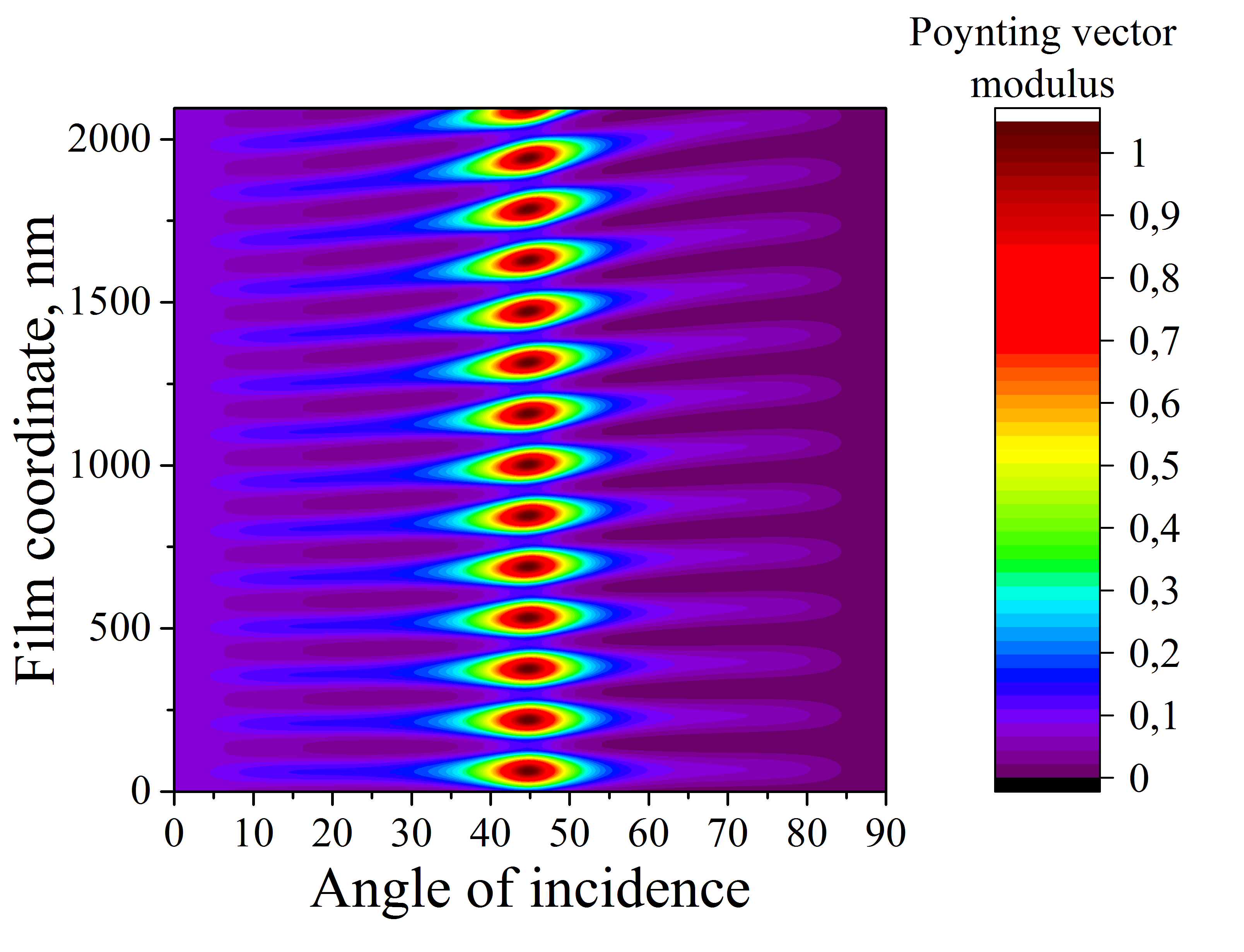}
	\caption{Modulus of Poynting vector in the $CeO_2$ film cross section for coherent s-polarized wave. The horizontal axis indicates the angle of wave incidence, the vertical axis indicates the distance to Al sublayer. The thickness of the $CeO_2$ film - 2100~nm, the dielectric permittivity at the wavelength of 632.8~nm $\varepsilon_{CeO_2}=4.595$ ($n_{CeO_2}=2.1436$). The parameters of the $Al$ sublayer: $d_{Al}=200$~nm, $\varepsilon_{Al}=-51.4+18.4i$~\cite{Rakic}. The dielectric permittivity of the substrate is $\varepsilon_{Al_2O_3}=2.89$ ($n_{Al_2O_3}=1.70$). Poynting vector of the incident wave is normalized to 1.\label{Poynting}}
\end{figure}

The excitation of the RRM by the incident wave leads to enhancement of the electric field inside the $CeO_2$ layer (see Fig.~\ref{Poynting}). As a result, the dipole moment representing the inhomogeneity increases as soon as its value is directly proportional to the electric field. Hence, the scattered field increases. At the same time, light absorption increases due to the electromagnetic energy concentration inside the system.

The considered model reflects the effect of increasing the scattering observed in the real system. When the coherent laser beam illuminates the sample at the angle corresponding to the excitation of an RRM, a considerable part of the beam energy is transferred to the mode of the system (see Fig.~\ref{Poynting}). The radiation mode propagates inside the film, where its energy is dissipating due to inhomogeneities and absorbed due to Ohmic losses in aluminum. Thus, the coherent radiation is transformed into the radiating dipoles through the RRM, and the scattering increases sharply (see Fig.!\ref{WeakScattering}). At the same time, the absorption coefficient also increases.

A similar effect was observed for $air/Ag/MgF_2/Ag$ layered system~\cite{SurfaceGarciaLlamos}, where it was shown that at the angles corresponding to the excitation of the guided modes, light scattering by the surface roughness increases. Thus, the scattering amplification effect is observed when either a guided mode or a resonant radiation mode (a Fabry-Perot resonator mode) is excited

To summarize, it can be concluded that the excitation of the waveguide RRM at a certain angle of incidence leads to the concentration of the incident wave radiation inside the film, which greatly increases both the scattering associated with the presence of inhomogeneities and the absorption due to Ohmic losses in aluminum. Therefore, at certain angles of incidence corresponding to the resonant radiation modes, both scattering and absorption coefficient maxima can be observed.

\subsection{Strong scattering, comparison with the experiment}
As it was mentioned earlier, the $CeO_2$ film possesses the inhomogeneous structure. The film contains various types of defects (gaps separating the facets of $CeO_2$, surface roughness of the film, surface roughness of the $Al$ sublayer, volume density inhomogeneities) acting as light scattering centers. Coherent radiation scattering by inhomogeneities of different types results in the transfer of its energy to that of the incoherent radiation. If the scattering is sufficiently large (the considerable part of the energy is transferred), it is necessary to solve the self-consistent problem; i.e. the coherent radiation energy losses due to the scattering should be taken into account. To formulate the self-consistent problem, we introduce the effective absorption of the coherent radiation in the $CeO_2$ film as an additional imaginary part of the $CeO_2$ effective refractive index
\begin{equation}
n^{eff}_{CeO_2}=n_{CeO_2}+in_{sc}.\label{Neff}
\end{equation}

The introduced value $n_{sc}$ should describe the transfer of the coherent radiation into the incoherent component. Its value depends on the concentration and scattering cross-section of all the inhomogeneities located inside the film. To clarify the meaning of $n_{sc}$, consider the propagation of a coherent radiation beam through a homogeneous medium containing scatterers. The intensity attenuation of the coherent radiation after passage of the section $dx$ can be described by the approximated transport equation~\cite{Ishimaru}
\begin{equation}
I(x+dx)=I(x)-I(x)\sigma Ndx\label{Transport}
\end{equation}
where $\sigma$ is the scattering cross-section of a single scatterer, and $N$ is the scatterer concentration. Thus, the intensity of the coherent radiation decays exponentially with the propagation length
\begin{equation}
I(x)=I_0e^{-\sigma Nx}.\label{Decay}
\end{equation}
In the case of the presence of different types of scatterers, Equation~(\ref{Transport}) turns into
\begin{equation}
I(x+dx)=I(x)-\sum_{i}I(x)\sigma_i N_idx\label{TransportMulti}
\end{equation}
where $\sigma_i$ and $N_i$ are the scattering cross-section and the concentration of the $i$-th inhomogeneity type. On the basis of Equation~(\ref{TransportMulti}), the single effective parameter can be introduced as
\begin{equation}
\sigma_{eff}N_{eff}=\sum_{i}I(x)\sigma_i N_i\label{Sigma}
\end{equation}
i.e. the problem with several types of scatterers can be reduced to the problem with one parameter. This fact is a result of the additive contributions of different scatterers to the coherent radiation intensity attenuation~(\ref{TransportMulti}).

At the same time, if a material has a complex refractive index $n=n_1+in_2$ (where the imaginary part is responsible for the effective absorption of the coherent radiation), the beam intensity attenuation in it can be described as
\begin{equation}
I(x)=I_0e^{-2n_2k_0x}\label{DecayN}
\end{equation}
where $k_0$ is the wavenumber in free space. The~(\ref{Decay}),~(\ref{Sigma}) and~(\ref{DecayN}) result in the estimation
\begin{equation}
n_2=\frac{1}{2k_0}\sigma_{eff}N_{eff}\label{NSigma}
\end{equation}
which connects the imaginary part of the effective refractive index with the scattering characteristics. Thus, the single parameter $\sigma_{eff}N_{eff}$ can be used for the description of the scattering. In this paper the value of this parameter is estimated based on the experimental data.

The estimation of the $CeO_2$ film parameters is performed by the optimization procedure for the film thickness and the effective refractive index. As an optimized function, a discrepancy between the experimental and theoretical reflectance curves of s-polarized light is chosen. The discrepancy is calculated by the $L_2$ norm. The Nelder-Mead method is used for optimization~\cite{NelderMead}. For the effective parameter $\sigma_{eff}N_{eff}$, the value $n_2=\frac{1}{2k_0}\sigma_{eff}N_{eff}\approx 0.0105$ is numerically obtained.

The optimized parameters of the $CeO_2$ film were used in calculations. It should be noted that the imaginary part of the effective refractive index, in this case, is not responsible for absorption (absorption occurs only in the aluminum layer) and rather describes the transformation of coherent radiation to incoherent radiation.

\begin{figure}
	\includegraphics[width=1\linewidth]{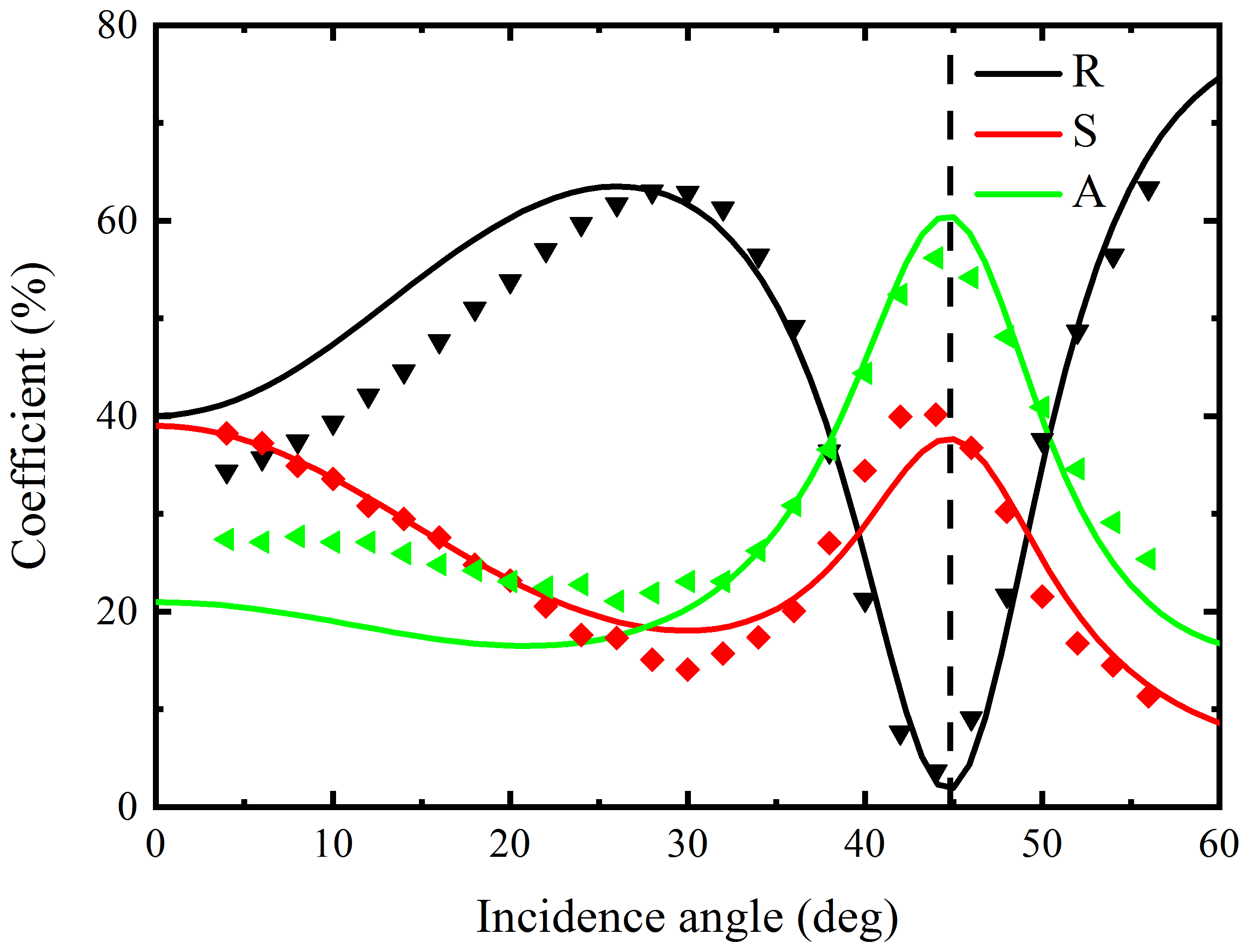}
	\caption{Dependences of the specular reflection, total scattering and absorption coefficients on the incidence angle of s-polarized radiation. Symbols indicate the values obtained experimentally. The calculated thickness of the $CeO_2$ film - 2096~nm, the refractive index at the wavelength of 632.8~nm $n_{CeO_2}=2.1436+0.0105i$. The parameters of the $Al$ sublayer: $d_{Al}=200$~nm, $\varepsilon_{Al}=-51.4+18.4i$~\cite{Rakic}. The dielectric permittivity of the substrate is $\varepsilon_{Al_2O_3}=2.89$ ($n_{Al_2O_3}=1.70$). The vertical dashed line indicates the angle of mode excitation (see Equation 2), and the dotted line corresponds to the Brewster angle.\label{Full}}
\end{figure}

The calculated reflectance, absorption and scattering coefficients as functions of the s-polarized laser beam incidence angle are shown in Figure~\ref{Full}. The reflectance $R$ is calculated using the T-matrix method. According to the energy conservation law, the sum of the scattering ($S$) and absorption ($A$) coefficients is
\begin{equation}
S+A=1-R.\label{ConservationLawFull}
\end{equation}

Both the absorption in aluminum and the scattering in cerium dioxide lead to the losses in coherent radiation energy. The coefficient $S$ is normalized to the intensity of the experimental value of the scattered radiation at the normal incidence. The remaining fraction of the energy is absorbed inside the aluminum.

The angular dependencies (see Fig.~\ref{Full}) calculated in the framework of the considered simple model are in good agreement with the experimental result. The dashed line indicating the Fabry-Perot mode (RRM) excitation angle coincides with the minimum of the reflection curve. That confirms the strong dependence of scattering and absorption coefficients maxima on the Fabry-Perot resonant modes. The revealed nature of scattering and absorption coefficients maxima opens up great prospects for manipulation of the reflection from the inhomogeneous films

\section{Conclusion}
The scattering of the coherent electromagnetic radiation by inhomogeneities of the $CeO_2$ film is considered in the paper. The electron-beam-deposited cerium dioxide is a material possessing inhomogeneities of different types. It is experimentally shown that the light scattering and absorption coefficients simultaneously increase at certain radiation incidence angles.

The scattering is theoretically studied in terms of the single-scattering approximation. The proposed theoretical model showed excellent agreement with the experimental data. The important point of the consideration is the use of a single adjustable parameter, namely, the effective scattering cross-section, in the model.

It is shown that the increase in scattering and absorption of the incident coherent wave occurs at the angles of incidence, at which the resonant radiation modes (the modes of a Fabry-Perot resonator) are excited. Thus, the increase in scattering and absorption occurs due to the field enhancement inside the film by the resonant radiation modes of the system. The scattering of coherent electromagnetic radiation by inhomogeneities of cerium oxide film is considered in the paper. The electron-beam deposited cerium dioxide is a material possessing inhomogeneities of different types. It is experimentally shown that the scattering and absorption of light simultaneously increase at certain radiation incidence angles.

\section{Acknowledgment}
This work was supported by the Presidium of the Russian Academy of Sciences (the Basic Research Program I.40 "Creation of ultra-sensitive methods for identification of biological objects through optical metamaterials") and by Russian Foundation for Basic Research (RFBR) (grant 18-52-00044).

\section{Appendix}
\begin{figure}
	\includegraphics[width=1\linewidth]{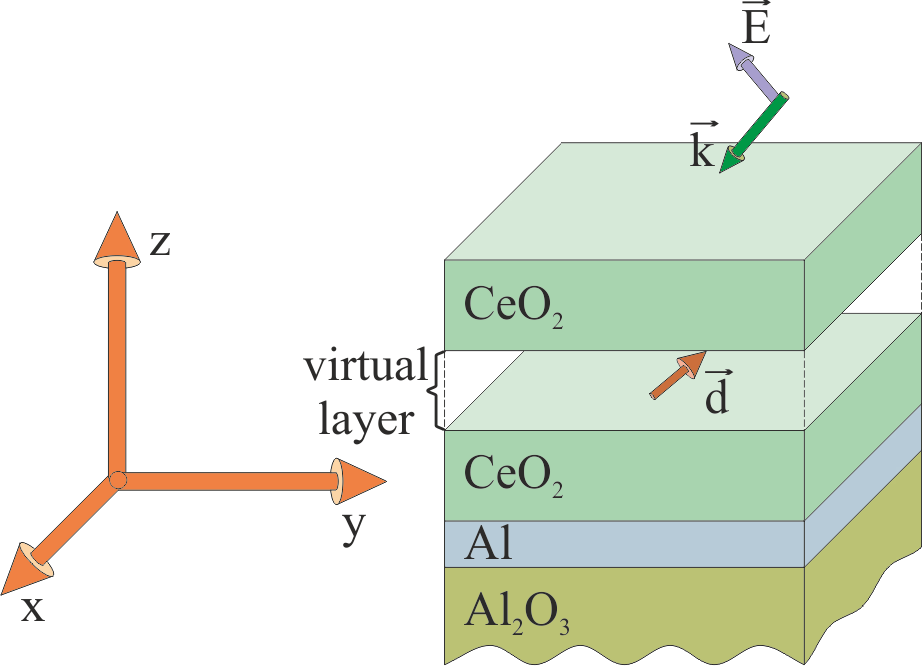}
	\caption{Model of a system with a dipole scatterer.\label{Dipole1}}
\end{figure}

We consider a single dipole inside the layer of $CeO_2$ (see Fig.~\ref{Dipole1}). The magnitude of the dipole will be assumed to be linearly dependent on the magnitude of the field
\begin{equation}
\vec{d}=\alpha\vec{E}(x,y,z)\label{linearD}
\end{equation}
where $\alpha$ is the inhomogeneity polarizability. In other words, in our model the orientation and magnitude of the dipoles are determined by the distribution of the coherent electric field in the sample. The magnitude of the electric field of the coherent radiation (see Fig.~\ref{Dipole1}) will be calculated by using the T-matrix method~\cite{Yeh}.

In the statement of the scattering problem given in the paper, the calculation of the scattered light amplitudes is equivalent to the calculation of the dipole-induced field. The following approach to the field calculation is equivalent to calculating the Green function for a dipole placed into the structure under consideration.

The dipole acts as a radiating source. Let us consider the radiation of a dipole in a vacuum. We will consider the coordinate system depicted in Fig.~\ref{Dipole1}. The polarization vectors are
\begin{equation}
\vec{e}_s=\begin{pmatrix}
\frac{k_y}{k_\parallel} &
-\frac{k_x}{k_\parallel} &
0
\end{pmatrix}^T\label{Es}
\end{equation}
for s-polarization and
\begin{equation}
\vec{e}_p=\begin{pmatrix}
\frac{k_xk_z}{kk_\parallel} &
\frac{k_yk_z}{kk_\parallel} &
-\frac{k_\parallel}{k}
\end{pmatrix}^T\label{Ep}
\end{equation}
for p-polarization. Here $k$ is the wavenumber in vacuum, and $k_\parallel=\sqrt{k_x^2+k_y^2}$. The electric dipole field can be written in the form~\cite{ChenToTai}
\begin{equation}
\vec{E}=ik\left(\vec{A}+\frac{1}{k^2}\vec{\bigtriangledown}\left(\vec{\bigtriangledown},\vec{A}\right)\right)\label{Field}
\end{equation}
where $\vec{A}$ is the vector potential, equal to
\begin{equation}
\vec{A}(\vec{r})=-\frac{i\omega}{c}\frac{\exp^{ik|\vec{r}-\vec{r}_0|}}{|\vec{r}-\vec{r}_0|}\vec{d}_0\label{Apotential}
\end{equation}
Here, $\vec{r}_0$ is the radius vector of the dipole position, and $\vec{d}_0$ is the dipole moment magnitude. Applying Weyl representation for a spherical wave~\cite{Mandel}
\begin{equation}
\frac{e^{ikr}}{r}=\frac{i}{2\pi}\iint_{-\infty}^{+\infty}\frac{e^{i(k_xx+k_yy+k_zz)}}{k_z}dk_xdk_y\label{Weyl}
\end{equation}
one obtains the expansion over the basis of the plane waves
\begin{multline}
\vec{E}=\frac{i}{2\pi}\iint_{-\infty}^{+\infty}\frac{1}{k_z}
\begin{pmatrix}
k^2-k_x^2 & -k_xk_y & -k_xk_z\\
-k_xk_y & k^2-k_y^2 & -k_yk_z\\
-k_xk_z & -k_yk_z & k^2-k_z^2
\end{pmatrix}
\begin{pmatrix}
d_x\\d_y\\d_z
\end{pmatrix}
\\
\cdot{e^{i(k_xx+k_yy+k_zz)}}dk_xdk_y\label{Eexpand1}
\end{multline}

The expression~(\ref{Eexpand1}) for the electric field can be decomposed over the plane waves of different polarizations
\begin{multline}
\vec{E}=\frac{i}{2\pi}\iint_{-\infty}^{+\infty}\left(E_s(k_x,k_y)\vec{e}_s+E_p(k_x,k_y)\vec{e}_p\right)\\\cdot e^{i(k_xx+k_yy+k_zz)}dk_xdk_y\label{Eexpand2}
\end{multline}
where
\begin{equation}
E_s(k_x,k_y)=\frac{k^2}{k_\parallel k_z}(k_yd_x-k_xd_y)\label{FieldS}
\end{equation}
for s-polarization and
\begin{equation}
E_p(k_x,k_y)=\frac{k}{k_\parallel}(k_xd_x+k_yd_y)-\frac{kk_\parallel}{k_z}d_z\label{FieldP}
\end{equation}
for p-polarization.
\begin{figure}
	\includegraphics[width=1\linewidth]{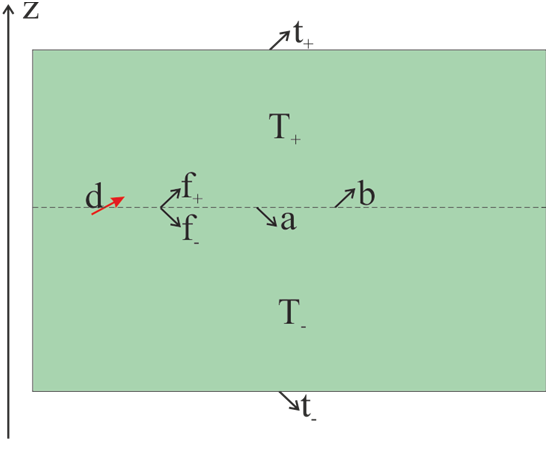}
	\caption{A dipole placed between two layered subsystems.\label{Dipole2}}
\end{figure}

It is further convenient to calculate the dipole-induced field inside the structure by using the T-matrix method.

It is easier to carry out the calculation by distinguishing two parts of the system: subsystems above and below the dipole (see Fig.~\ref{Dipole2}). Then, the dipole is placed inside an infinitely thin layer separating the subsystems. In this case, the radiation of a dipole inside this infinitely thin layer can be described by formulas~(\ref{Eexpand2}),~(\ref{FieldS}),~(\ref{FieldP}). The problem of electromagnetic wave propagation can be calculated with the help of the T-matrices $T_+$ and $T_-$, respectively (see Fig.~\ref{Dipole2}). The field at the upper (lower) boundary of the system can be found by solving the following system of equations:
\begin{subequations}
	\label{Tmatrix}
	\begin{eqnarray}
	\begin{pmatrix}
	t_+\\0
	\end{pmatrix}
	=
	T_+
	\begin{pmatrix}
	f_++b\\a
	\end{pmatrix}
	\\
	\begin{pmatrix}
	b\\a+f_-
	\end{pmatrix}
	=
	T_-
	\begin{pmatrix}
	0\\t_-
	\end{pmatrix}
	\end{eqnarray}
\end{subequations}
where the amplitudes $t_+$, $t_-$, $a$, $b$, $f_+$, $f_-$ of the waves are shown in Figure~\ref{Dipole2}. The amplitudes $a$ and $b$ correspond to the waves reflected from the lower and upper subsystems, and $f_+$ and $f_-$ are the amplitudes of the waves induced by the dipole (see~(\ref{Eexpand2}),~(\ref{FieldS}),~(\ref{FieldP})). Knowledge of the wave amplitude $E^+_{s,p}$ on the upper plane of the system allows one to obtain easily the field at infinite distance according to the formula
\begin{equation}
E^{(\infty)}_{s,p}(\frac{x}{r},\frac{y}{r},\frac{z}{r})=-ik_zE^+_{s,p}(k_x,k_y)\frac{e^{ikr}}{r},r\to\infty
\end{equation}
where $E^{(\infty)}_{s,p}(\frac{x}{r},\frac{y}{r},\frac{z}{r})$ is the amplitude of polarized light at a large distance in the direction of the vector $(\frac{x}{r},\frac{y}{r},\frac{z}{r})$. Thus, the scattering amplitude can be calculated. The integrating of the scattering amplitude modulus square over the hemisphere is the hemispherical scattering.
% Create the reference section using BibTeX:
\bibliography{scattering1}

\end{document}